\documentclass[prb,preprint]{revtex4-1} 
\usepackage{amsmath}  
\usepackage{amsfonts} 
\usepackage{multirow}
\usepackage{graphicx} 
\usepackage[utf8]{inputenc}
\usepackage{braket}
\usepackage{longtable}
\usepackage{url}
\usepackage{xcolor}

\begin{document}
\title{A Hands-On Quantum Cryptography Workshop For Pre-University Students}
\author{Adrian Nugraha Utama}
\email{adrian\_utama@u.nus.edu} 
\author{Jianwei~Lee}
\author{Mathias Alexander Seidler}
\affiliation{Centre for Quantum Technologies, National University of Singapore, Block S15, 3 Science Drive 2, 117543}

\date{\today}

\begin{abstract}
We developed a modified version of a conventional (BB84) quantum key distribution protocol that can be understood and implemented by students at a pre-university level. 
We intentionally introduce a subtle but critical simplification to the original protocol, allowing the experiment to be assembled at the skill level appropriate for the students, at the cost of creating a security loophole.
The security vulnerability is then exploited by student hackers, allowing the participants to think deeper about the underlying physics that makes the protocol secure in its original form.
\end{abstract}

\maketitle 

\section{Introduction}
Quantum cryptography promises information-theoretic security based on quantum mechanics.~\cite{qkdreview} 
Secure communication between two parties typically consists of two steps. First, a key is distributed between two parties. Second, parties use the key to encrypt their messages. 
Classical cryptography~\cite{rsa} relies on either ``safely" distributed symmetric keys (e.g. via smart cards), or computational complexity (e.g. factorization), and requires trusting the courier, or can be compromised with either increased computational power 
or with quantum algorithms.~\cite{niel} 
However, quantum cryptography does not have these shortcomings, as it provides a framework, based on quantum mechanics, to determine if a would-be adversary has gained enough information to compromise the secrecy of distributed keys.~\cite{bb84proof, bb84proof2}

In 1984, Bennett and Brassard invented a scheme~\cite{bb84} to distribute symmetric keys securely between two parties. The security is based on the no-cloning theorem:~\cite{nocloning, eprdevices} an eavesdropper cannot create identical copies of an arbitrary unknown quantum state. This quantum key distribution (QKD) protocol is known as the BB84 protocol.
In 1991, Artur Ekert introduced another important QKD scheme~\cite{ekert} which utilizes entanglement~\cite{entg} 
and a Bell violation measurement~\cite{bell} to guarantee privacy.
These inventions spurred considerable interest from both the computer science and physics communities, and QKD is now an emerging technology.~\cite{exp1, exp2, exp3} Based on counter-intuitive, quantum-physical concepts, QKD also intrigues the general public. 
There now exist several QKD demonstrations for non-experts which focus on teaching the underlying physics of QKD.~\cite{momentum, brazildemo, choco, entangleme, interactivesims, interactivetuts}
However, a full demonstration that involves both the use of QKD to generate the encryption key, and subsequently applying the key for encrypting a secret message, will help learners to appreciate how both relatively non-trivial procedures are integrated to realize a fully-functional quantum encryption system.

In this work, we present a quantum cryptography workshop developed for pre-university students (age 15 to 19). 
Our aim is to go from abstract to concrete; 
from understanding quantum-mechanical concepts, 
to implementing a working setup in the laboratory.
First, the students set up two communication channels: a classical channel based on infrared pulses, and a quantum channel based on polarization encoded photons. 
Second, they distribute a symmetric key between two parties with the BB84 protocol.~\cite{bb84} 
Finally, they use the key to encrypt secret messages and send them over to the other party via the classical channel. 


The security of BB84 relies on the fact that a single quantum bit (qubit) cannot be copied.~\cite{bb84proof} 
When multiple qubits of the same state, e.g. multiple photons with the same polarization, are distributed, security is compromised
since a fraction of the qubits can be intercepted and measured by an adversary (a form of side-channel attack~\cite{diamanti2016practical}).
In this workshop, we demonstrate this attack by using macroscopic laser intensities consisting of millions of photonic qubits per coherence time, creating an exploitable security loophole. 
This security loophole can be addressed by attenuating macroscopic sources to a mean photon number well below one per coherence time.~\cite{bennett1992experimental} 
In commercial implementations, more sophisticated decoy-state protocols have been adopted to allow higher photon numbers per pulse -- useful for transmitting over long distances.~\cite{Hwang2003decoystate}
However, we intentionally leave this loophole open to allow the students to revisit the `no-cloning' theorem and its role in quantum cryptography -- we task one group of students to implement a side-channel attack, retrieve the key, and decode secret messages.

In subsequent sections, we describe the BB84 protocol and our modification which allows it to be implemented with commercial, off-the-shelf components. 
To facilitate educators in conducting the workshop, we have provided detailed schematics and programs used to control the experiment.
Our approach received good feedback from the students, who appreciated the hands-on nature of the learning experience, and the unexpected security loophole present in what they presumed to be a secure quantum key distribution system.

\section{BB84 Protocol With a Twist}\label{sec:bb84}

\begin{figure}[ht!]
\includegraphics[width=8.6 cm]{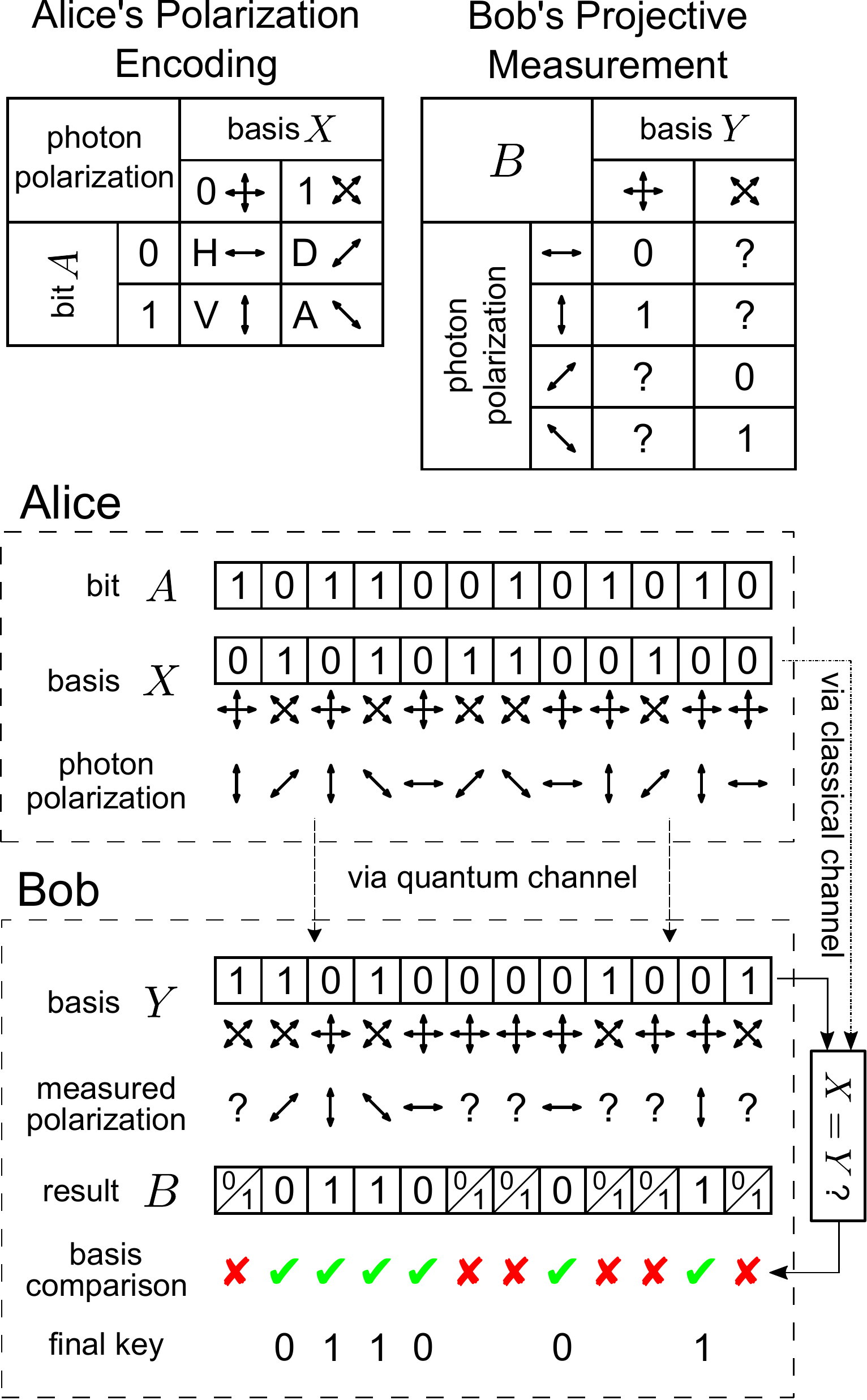}
\caption{\label{figure:bb84}
    Summary of the BB84 protocol.
    Alice generates a random sequence $A$ of bits to send over the quantum channel. For each bit, she chooses a basis $X$ to transmit a photon with the corresponding polarization (figure top-left) to Bob. 
    For every photon Bob receives, he randomly chooses a basis $Y$ to measure the photon's polarization (figure top-right). 
	After transmitting the entire sequence $A$, 
    Alice sends her basis sequence $X$ to Bob using the classical channel. 
    If Bob chooses the same basis as Alice for a particular photon ($X = Y$), Bob should measure the same polarization and he can accurately infer the value of Alice's bit.
    Alice and Bob keep only the bits corresponding to photons that were prepared and measured with the same basis, resulting in an identical bit sequence for both parties that they can utilize for encryption.
} 
\end{figure}

In this section, we describe our implementation of the BB84 protocol and its security assumption.
The protocol generates a symmetric encryption key between two parties, usually called Alice and Bob.

\begin{enumerate}
    \item Alice's part of the protocol
    
Alice generates a sequence of random bits $A = \{A_1, A_2, ..., A_n\}$, where $A_i\in[0,1]$ is the~$i\text{-th}$ bit. 
Next, she generates another random sequence $X = \{X_1, X_2, ..., X_n\}$, where $X_i\in[0,1]$ denotes the polarization basis used to encode $A_i$.
According to $A$ and $X$, Alice transmits a sequence of polarized photon pulses to Bob.
The encoding scheme is tabulated in Fig.~\ref{figure:bb84}. 

\item Bob's part of the protocol

Similar to Alice, Bob generates a random sequence of polarization measurement bases $Y = \{Y_1, Y_2, ..., Y_n\}$. 
He projects the incoming polarization according to $Y$, 
and measures the corresponding light intensity. 
The intensity values are categorized into low and high values and recorded as $B = \{B_1, B_2, ..., B_n\}$, with $B_i\in[0,1]$. 

\item Establishing the final key

Thereafter, Alice and Bob communicate their basis choices over a classical channel. 
When $X_i\neq Y_i$, $B_i$ does not provide any information about $A_i$, since the measurement bases are mutually unbiased and thus $P(A_i|B_i) = 0.5$ for any values of $A_i$ and $B_i$ - these results are discarded. 
Alice and Bob keep only outcomes when their basis choice agree to generate a shared secret key. 
This procedure is known as ``key sifting''.
\end{enumerate}

In the original BB84, each bit is encoded by a single photon. 
Security is ensured by two quantum mechanical concepts. 
First, to learn about the state transmitted from Alice, 
Eve would have to intercept the photon, 
and perform a measurement which inadvertently disturbs its state. 
Second, if Eve resends the photon she measured to Bob, 
she cannot create a perfect copy due to the quantum no-cloning theorem. 
Consequently, Alice and Bob will be able to identify the presence of Eve by checking for inconsistencies for a subset of the transmitted bits that were prepared and measured in the same basis. 
The rest of the bits are then used to generate the final encryption key.

We guide students to build a working implementation of the protocol -- with a twist, which breaks the security of the protocol. We use laser pulses containing many copies of the same quantum state, which allows for a side-channel attack: An eavesdropper (Eve) can analyse a fraction of the transmitted signal and get full information about the state. In the subsequent sections, we refer to these laser pulses as ``qubits'', with the quotation mark, to emphasize their simulated property.

\begin{figure}[ht!]
\includegraphics[width=8.6cm]{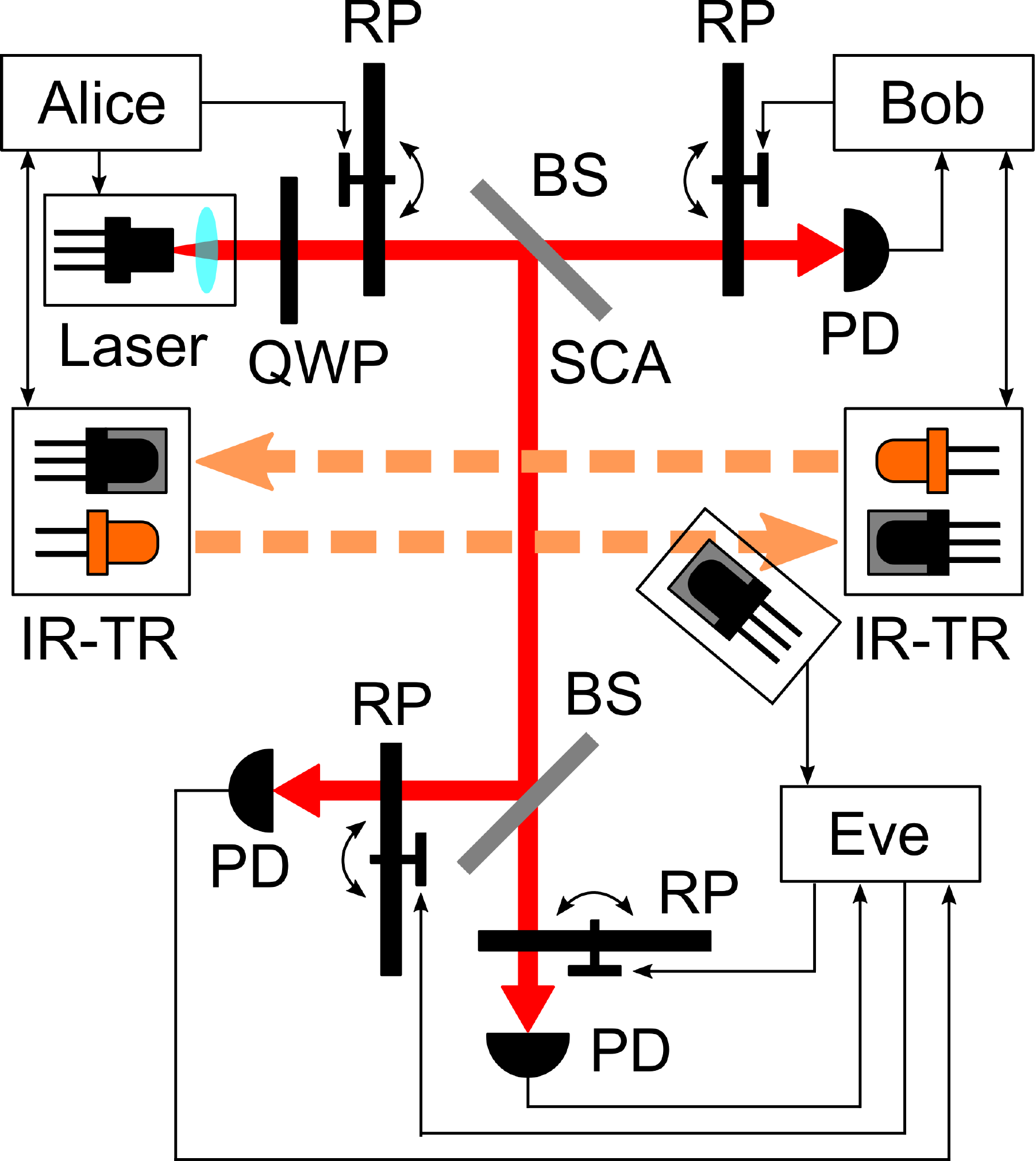}
\caption{\label{figure:setup}
BB84 setup (quantum channel): Alice encodes a string of bits using different linear polarization choices -- created using a pulsed laser source, quarter-wave plate (QWP) and a rotatable polarizer (RP).
Bob projects Alice's photons into different polarization bases, and measures the corresponding intensity with a photodetector (PD).\newline
Classical channel: Using infrared transceivers (IR-TR), Alice and Bob communicate the matched bases and the encrypted message.\newline
Side-channel attack (SCA): 
Using a beam splitter (BS), Eve splits off Alice's photons and measures them in two different bases simultaneously. 
She also intercepts the matched bases and encrypted message using her IR receiver from the classical channel. 
} 
\end{figure}

\begin{figure}[t]
\includegraphics[width=8.6cm]{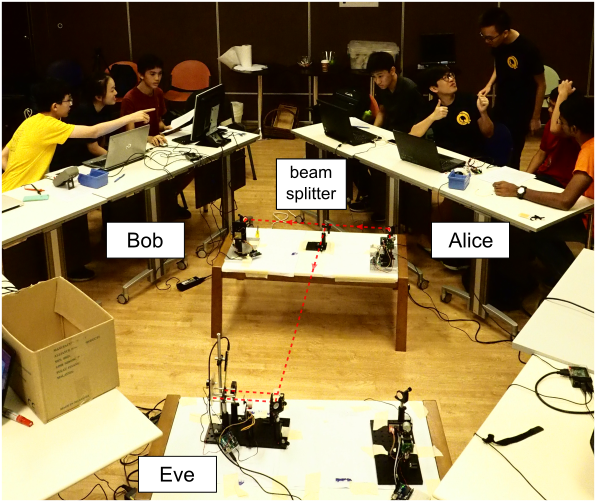}
\caption{\label{figure:student_setup}
 BB84 setup as implemented in the workshop. Optical components for the quantum channel were placed and aligned on small optical breadboards on top of short wooden tables -- the beam height was below eye level to ensure safety. Students are seated in a diamond configuration enclosing the setup, preventing accidental misalignment. The dashed line indicates the laser beam path.
} 
\end{figure}

\section{\textbf{Experimental Implementation}}\label{sec:implementation}
We design a workshop for students to gain insight to the inner workings of the protocol by tasking them to construct and operate the modified version of the protocol, described in the previous section.
Students establish the quantum channel using polarization-encoded ``qubits'' by controlling laser diode systems, motorized polarization analyzers, and photodetection circuits.
They also establish a public classical channel using infrared (IR) pulses by assembling IR transmitting and receiving circuits. 
A separate group of students eavesdrop on these channels using a beamsplitter and additional IR devices.
Figure~\ref{figure:setup}~and~\ref{figure:student_setup} shows the schematic for the entire setup and the classroom arrangement.

In the following, we describe the procedure for establishing the quantum and classical channels. The experimental sequence was largely automated as it requires synchronization between different elements, as well as for maintaining consistency in the generation of every bit. The sequence was executed using Arduino and Python programs. These are fully documented and available on Github.~\cite{github}

\subsection{Quantum Channel}
\begin{figure}[ht!]
\includegraphics[width=17.2cm]{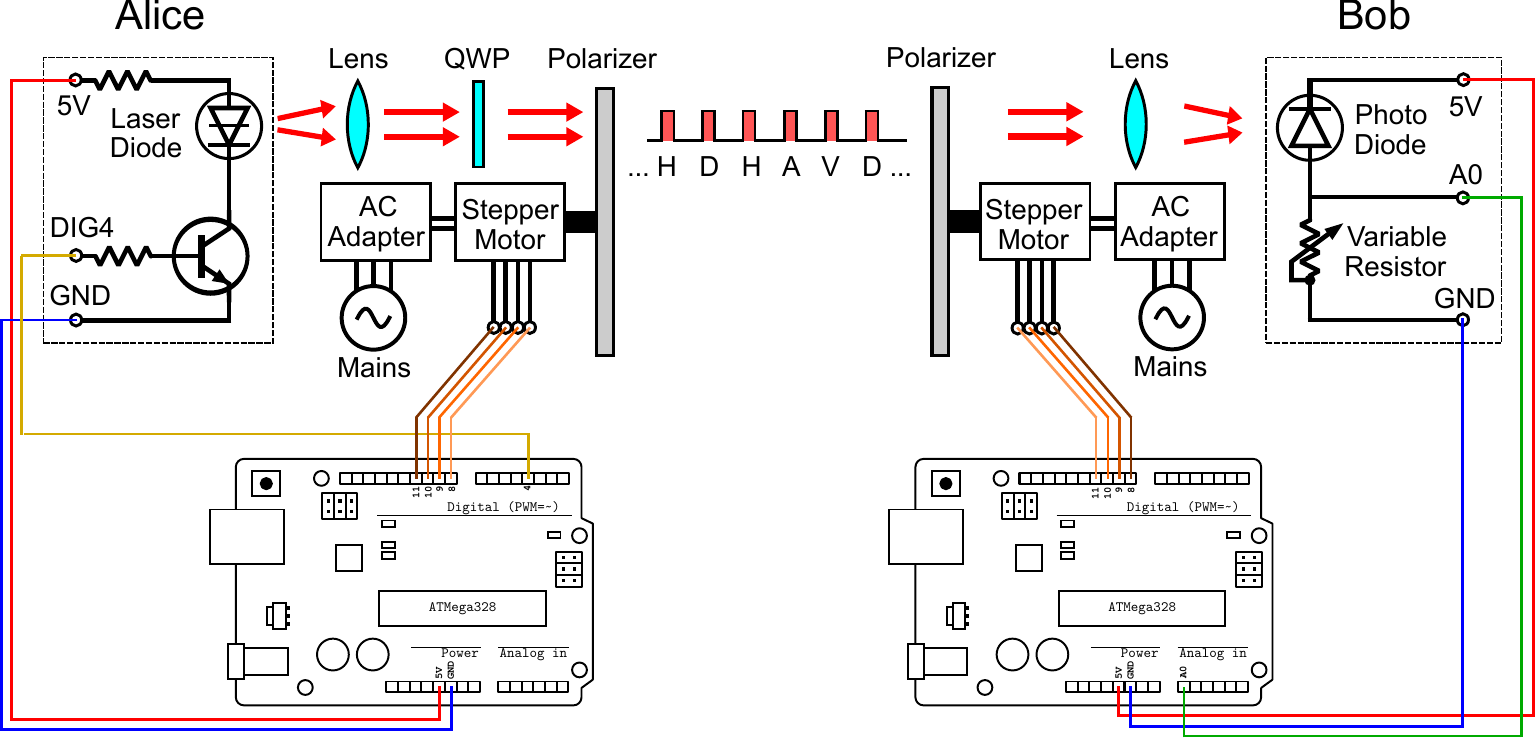}
\caption{\label{figure:quantum_setup}
Schematic of the quantum channel. 
Alice transmits a series of polarized light pulses to Bob -- created using a laser diode, a quarter-wave plate (QWP) and a motorized polarizer. Bob projects the incoming states with his motorized polarizer and measures its intensity with a photodetector. 
The motors and laser diode are operated via the digital output on the microcontroller (DIG pins), while the photodiode readout is recorded by the analog-to-digital converter (ADC) on the microcontroller (A pins). QWP: Quarter-wave plate. GND: Ground connection.
} 
\end{figure}

\begin{figure}[ht!]
\includegraphics[width=8.6cm]{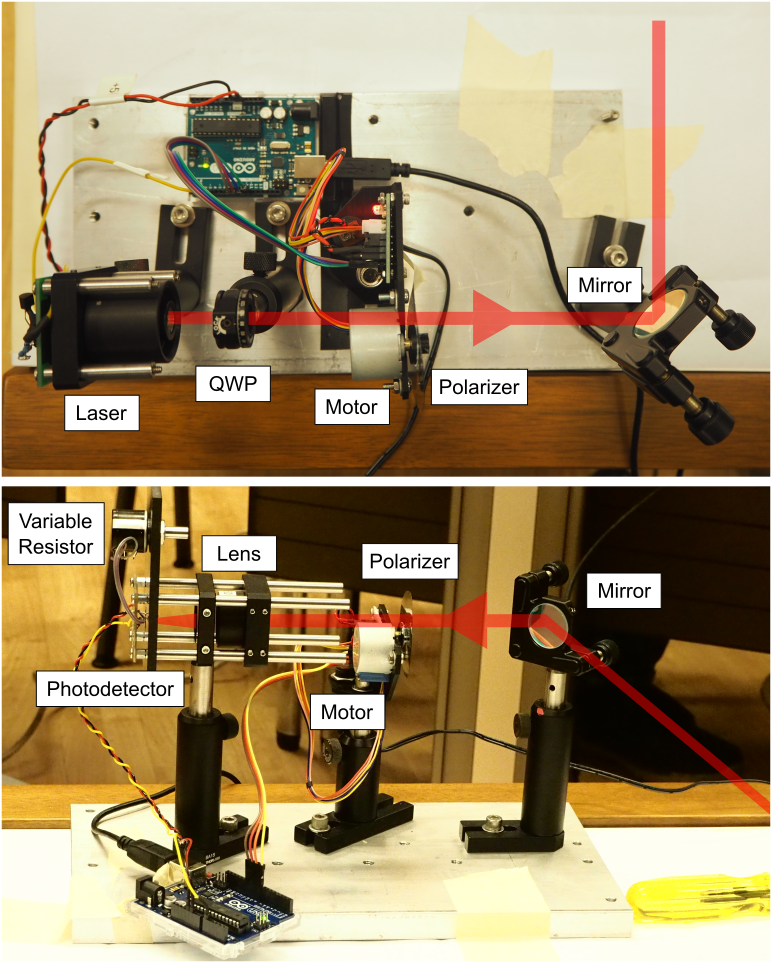}
\caption{\label{figure:quantum_pict}
 The quantum setup of Alice (top) and Bob (bottom). The laser beam path is provided as a guidance.
} 
\end{figure}

Figure~\ref{figure:quantum_setup}~and~\ref{figure:quantum_pict} illustrate the quantum channel.
Alice prepares different polarization states with laser diode pulses, a quarter-wave plate (QWP) and a motorized polarizer.~\cite{footnote1}
To perform a measurement on the incoming polarization state,
Bob chooses one of two measurement bases (H/V or D/A) at random, 
implements his choice with a motorized polarizer, 
and measures the corresponding light intensity. 

Before using the quantum channel,
both parties need to agree on a common coordinate system for their polarization states.
This is performed via a visibility measurement.~\cite{footnote2}
Alice begins her portion of the BB84 protocol by using an Arduino microcontroller to produce random~\cite{entropy} binary strings $A$ and $X$ 
-- the strings determine the sequence of polarization states to be transmitted.~\cite{footnote3} 
A random binary string $Y$ generated by Bob beforehand determines the bases used for projecting the transmitted ``qubits''. 
A time synchronization protocol ensures each ``qubit'' is delivered intact.~\cite{footnote4}
For each ``qubit'', the light intensity measured after the polarizer is compared with a predetermined threshold set at half the expected maximum light intensity. 
When Alice and Bob transmit and receive using the same basis, Bob observes two distinct intensity classes: high and low, allowing his measurements to be encoded as a binary number. 
Measuring the entire sequence generates a binary string $B$.
Intensities measured when both parties use different bases are discarded in subsequent steps.

\subsection{Classical Channel}\label{sec:classical}
\begin{figure}[ht!]
\includegraphics[width=8.6cm]{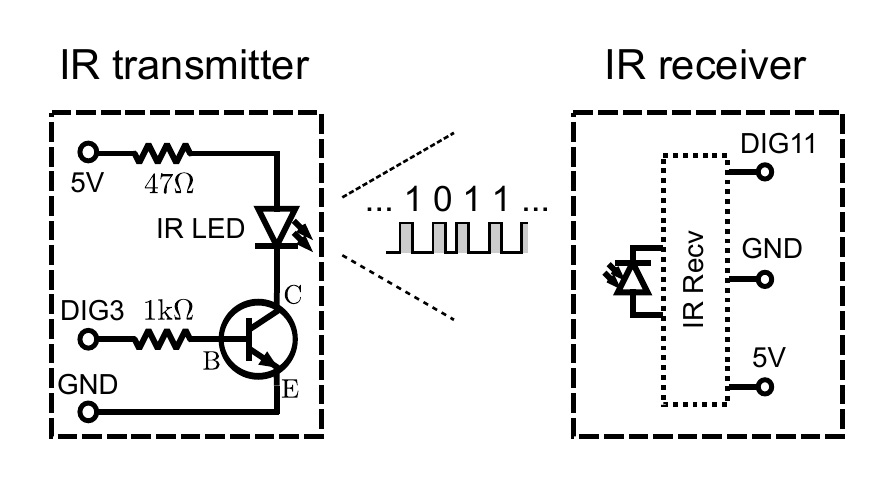}
\caption{\label{figure:classical_setup}
Schematic of the classical channel. 
A series of IR pulses enables Alice and Bob to communicate with each other. 
This enables them to exchange classical information e.g. basis choices, encrypted messages.
Information, encoded as binary strings, is translated into pulse sequences. 
A microcontroller uses the pulse sequences to switch the state of an IR LED to transmit the message.
An IR receiver detects the pulses and decodes it with a microcontroller.
} 
\end{figure}

\begin{figure}[ht!]
\includegraphics[width=8.6cm]{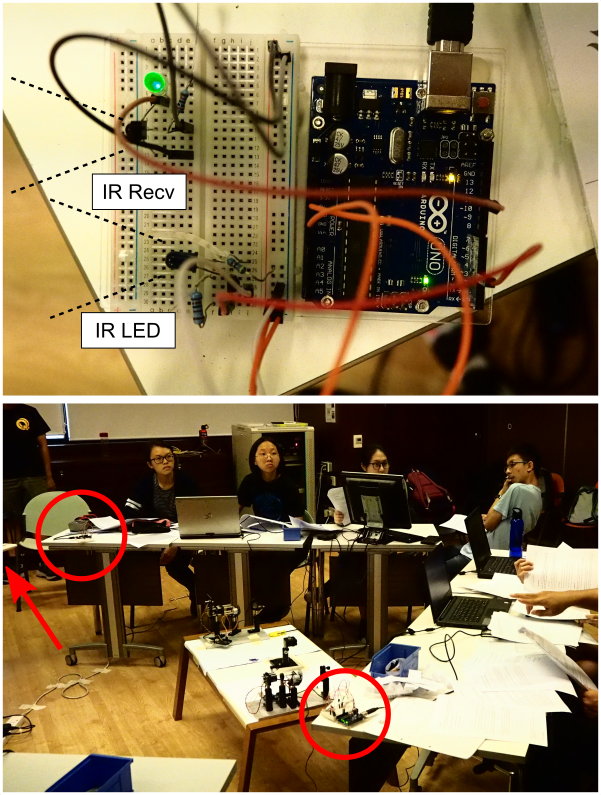}
\caption{\label{figure:classical_pict}
 Top: IR transmitter and receiver circuits are assembled on a solderless electrical breadboard. 
 Bottom: The circles indicate Alice and Bob's IR transceiver circuits, while the arrow indicates the location of Eve's IR receiver. The spatial mode of the IR LED has a relatively large solid angle.
} 
\end{figure}

The next step of the protocol is for Alice and Bob to communicate their basis choice over a  classical channel. 
In our implementation, the channel is established with IR transmitters and receivers (Fig.~\ref{figure:classical_setup} and Fig.~\ref{figure:classical_pict}).
Messages, encoded in a binary string sequence, are translated into IR pulsed sequences using the microcontroller.~\cite{irremote}
We chose to establish the classical channel with infrared LEDs due to their wide-spread availability in many consumer-electronic devices -- a hands-on experience assembling and operating the electronic circuits equips students to embark on their own projects after the workshop.

To generate a key common to both parties using $A$ and $B$,
Alice sends her basis choices $X$ to Bob, who compares it with his basis choices $Y$.
He then transmits the mask $X = Y$ back to Alice.
The mask corresponds to them transmitting and measuring in the same basis.
Applying the mask on the strings $A$ and $B$ generates the key $K$.
This step concludes the key distribution protocol.

Typically, a subset of K is checked for inconsistencies to reveal the presence of adversary Eve, when she performs an intercept-resend attack. 
As our implementation of Eve does not resend ``qubits'' to Bob, we omitted this step to simplify the setup.~\cite{footnote_extra}

\subsection{Building and applying the encryption key}\label{sec:encryption}
The length of $K$ is not fixed since the number of events where $X=Y$ is random. 
Thus, the key generation procedure is repeated until the desired key length (32 bits) is obtained.~\cite{footnote5}
The key can be used as a one-time pad to encode a message 32-bit long. 
This one-time pad guarantees the security of the message, provided that the key is only used once, and the key can not be obtained or guessed by an adversary.~\cite{bellovin}  
To encrypt a message $M$, Alice (or Bob) applies the bitwise exclusive or (XOR) operation $M \oplus K =: C$, 
obtaining the encrypted ``cipher'' $C$ of the original message that is to be transmitted through the classical channel.
To decode $C$ and recover $M$, the operation $C \oplus K$ is applied by the remote party.

Messages longer than 32 bit can still be encoded with a 32 bit key --
we used a protocol based on the XOR operation above (see Appendix A).
Although this results in insecure encryption,
it has the advantage of allowing students to encode longer messages without having to understand the workings of more sophisticated encryption techniques, which might detract from the main point of the workshop.

\subsection{Hacking the key distribution protocol}\label{sec:hack}
To eavesdrop on the quantum channel,
a third party (Eve) exploits the fact that macroscopic light pulses were used to represent ``qubits'', and directs part of the transmission in the channel into her own polarization analyzing setup (Fig.~\ref{figure:setup}) using a beam splitter.

As Eve's basis choice is \textit{a priori} not aligned to Alice and Bob's, she may not be able to distinguish between polarization states optimally.~\cite{footnote6}
However, by measuring in more than one basis choice simultaneously, she improves her ability to identify distinct polarization states even in the presence of laser intensity noise.~\cite{footnote7}

\begin{figure}[htb]
\includegraphics[width=8.6cm]{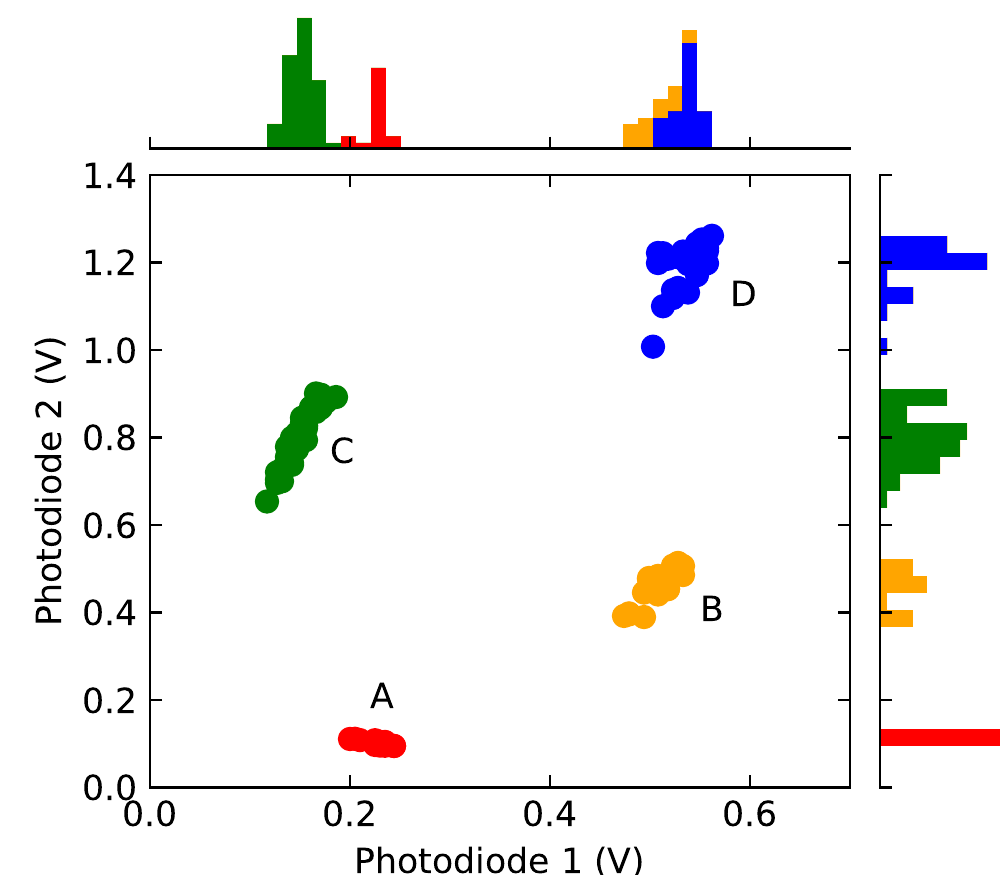}
\caption{\label{figure:kmeans}
    Four identified clusters of signal voltages measured by Eve's dual photodiodes 1 and 2 that measure light intensity after projecting the incoming polarization in two different basis (see Fig.~\ref{figure:setup}).
    Each cluster represents a polarization state intercepted in the quantum channel and are arbitrarily assigned A, B, C, D. 
    Each polarization state represents H, V, D or A sent from Alice to Bob. 
    By assigning the correct polarization state through trial and error, Eve is able to derive the transmitted key. 
    This is done via obtaining intelligible messages by decoding the ciphers transmitted through the classical channel.
} 
\end{figure}

In this way, each polarization state is associated with two intensity values (see Fig.~\ref{figure:kmeans}). 
In this 2-dimensional space, we use a K-means clustering algorithm to identify four distinct groups that correspond to the polarization states used.
The algorithm iteratively computes the positions of the four groups so as to minimize the least-squared distance between each data point and the mean position of its assigned group.~\cite{duda2012pattern}
Polarization states were arbitrarily assigned to each group and used to generate possible permutations of the key.

To proceed, Eve intercepts part of the light emitted from Alice's IR transmitter over the classical channel (Fig.~\ref{figure:classical_pict}).
For her first intercept, she obtains the bit positions corresponding to the sifted key (step 3 in Section~\ref{sec:bb84}) from the classical channel -- she uses this information to sift the permuted keys. 
Next, she intercepts cipher texts over the classical channel.
She uses them to identify the correct key permutation that results in a legible decoded message.


\section{The learning experience}
We aspire to create a learning experience where students are exposed to the inner-workings of current technologies with activities resembling hands-on lab experiences. This extension to a typical lab visit utilizes ideas and components already familiar to the students -- a platform to introduce the complex systems found in the typical research laboratories. 

We trialled this idea at a week-long outreach camp for pre-university students 
(aged 15-19) in Singapore,
where they learn about quantum technologies in a series of lectures and lab visits (QCamp).~\cite{qcamp} 
We also give every students the introductory book ``Six Quantum Pieces'',~\cite{sixquantum} which is an excellent resource on quantum physics aimed for high school students.

In the first iteration of the workshop, we tasked the students to custom-build an IR remote controller using the same know-how required for the classical channel (Section \ref{sec:classical}). As we were encouraged by positive responses from the students and organisers, we decided to extend the workshop into a full BB84 experiment. 

\subsection{Planning considerations}

We designed the implementation of the workshop to be simple and cost-effective.
This is addressed primarily by the use of macroscopic light intensities in the quantum channel with centimeter-sized, commercially available, Si-PIN photodiodes. 
This removes the requirement to focus light to millimeter-sized single-photon detectors (eg. avalanche photodiodes) usually required for BB84. 
To further reduce cost, we use standard off-the-shelf optical and mechanical components (Appendix B). 
Whereas the vulnerability created from using macroscopic light is undesirable in a security context, we found that the failure to realize a ``qubit'' as a single quanta provides a powerful learning opportunity for students who successfully hack the system (Section \ref{sec:hack}) despite following the key distribution procedure faithfully (Section \ref{sec:encryption}).

Another powerful motivation for simplifying the setup is to allow students the opportunity to build a working QKD system from the ground up -- a hands-on approach generally increases students’ engagement compared to didactic methods.~\cite{handson} A simple setup could be built within the time-constraints expected for an extra-curricular activity. Furthermore, our approach allows students to demonstrate competence, which is an essential intrinsic motivating factor in any learning task.~\cite{ryan}

Our approach complements previous quantum cryptography demonstrations realized by Lemelle et.al.~\cite{momentum} and Camargo et.al.,~\cite{brazildemo}
who also used macroscopic light pulses as a proxy for single photons to encode ``qubits''. 
While the effects on the transmitted ``qubits'' during an intercept-resend attack has been explored,~\cite{brazildemo}
we focus instead on providing the students with the hands-on experience of decrypting the intercepted photons. 
This we think, provides a visceral experience for the students who witness the retrieval of secret transmission using a purportedly quantum-secure protocol executed with classical resources. 

\subsection{Learning objectives}

Before the workshop, the students underwent a series of QCamp lectures covering a spectrum of basic quantum physics and technology discourses. A few lectures serve as the prerequisites to our workshop: photon polarisation, classical cryptography, and quantum key distribution (particularly BB84). The total time for these lectures are about 3 hours.

Although the workshop might appear to require a relatively advanced syllabus, we were able to keep the students focused on the main tasks to build up the cryptographic system. We leave it for the students to explore the more technical side of the protocol implementation on their own.~\cite{github}

We limited the learning objectives for the students in order to keep the workshop to a time limit of 3 hours: 
\begin{enumerate}
    \item Students should be able to assemble and operate experimental apparatus to distribute encrypted messages with keys obtained from BB84 -- refer to Section~\ref{sec:implementation} for the setup description and Section~\ref{sec:sturole} for the student's tasks. 
We use these processes as pedagogical tools to help students understand the cryptography protocol more concretely. Two examples are as follows:
(i) establishing a common polarization basis between Alice and Bob allows students to physically implement ``qubits'' in the polarization degree-of-freedom,
(ii) implementing the classical channel with IR pulses provides, for most students, a first encounter of how pulse sequences are used to transmit information.

    \item Students should be able to understand the central operating principle of QKD after the hacking attempt -- the no-cloning theorem. The apparent failure of the QKD system creates a cognitive conflict which has been shown to result in greater learning gains.~\cite{cogconflict, cogconflict2}
\end{enumerate}

\begin{figure}[ht!]
\includegraphics[width=8.6cm]{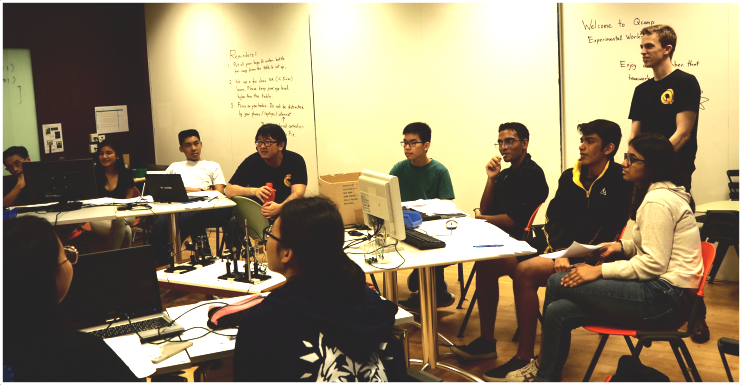}
\caption{\label{figure:debrief}
 Eve is revealed -- students are discussing in the debriefing session.
} 
\end{figure}

The workshop ends when Eve successfully decodes the secret message. The debriefing session (Fig.~\ref{figure:debrief}) usually starts with Eve reading out loud the secret message that Alice sends to Bob. This unexpected revelation sparks the discussion of how the setup differs from an ideal BB84 setup -- our setup does not use light at the single photon level. We further review the assumptions necessary for BB84, the importance of the no-cloning theorem in quantum cryptography, and discuss how device-independent QKD protocols can alleviate some of the assumptions.~\cite{diqkd, diqkd2} 
Facilitators may also take this opportunity to explore other ingenious hacks that exploit the vulnerabilities of practical QKD systems.~\cite{blackpaper}

\subsection{Student's roles}\label{sec:sturole}

The students are divided into four teams corresponding to Alice, Bob and Eve (two teams). 
More students were assigned to Eve as the duration of our workshop limited the tasks 
that could be accomplished by each team -- 
Eve had to acquire the same knowledge as Alice and Bob, in addition to hacking the protocol.
Eve comprised of two teams Eve-classical and Eve-quantum, each focusing on their respective channels. Each team consists of 3-5 members. 
We assigned a facilitator to each group to help and guide the students throughout the tasks. 

The tasks assigned to each team are described below. Teams Alice and Bob were assigned tasks 1-4, team Eve-classical were assigned tasks 1, 5 and 7, and team Eve-quantum were assigned tasks 2, 6, and 7.   
Students were given three hours to accomplish them.

\begin{enumerate}
    \item Classical channel. The students build the IR transceiver on the electrical breadboard, and run pre-written programs which allows them to exchange messages back and forth. 
    \item Quantum channel. The students align the polariser to establish a common polarisation basis, and perform a basic 16-bit BB84 protocol by exchanging a paper (simulating a public channel) back and forth. We have assembled and aligned the optical components prior to the workshop.
    \item Encryption key generation. The students run an automated BB84 protocol which utilises both the classical and quantum channel to generate a 32-bit key. 
    \item Secret message communication. The students use the 32-bit key to encrypt/decrypt the secret messages with the XOR operation (for 32-bit messages or shorter). Alternatively, they run a program which 
    encrypt/decrypt long messages with a key expansion procedure (see Appendix A). Alice sends the encrypted message through the classical channel.
    \item Classical channel hacking. The students position the IR transceiver at a convenient location to eavesdrop on Alice's messages -- in particular the matched basis and the encrypted secret messages. 
    \item Quantum channel hacking. The students record the light pulses split off from Alice's beam using the polarization analyzer, and cluster the polarizations accordingly. A beam-splitter has been placed on Alice's beam path prior to the workshop.
    \item Cracking the secret message. The students work together -- Eve-classical gives the matched basis to Eve-quantum to generate all the possible key combinations. They use the key combinations on the intercepted cipher text to decode a legible message. 
\end{enumerate}

At the end of the workshop, we allocate time for the students to exchange learning experiences with each other, in particular the set of tasks performed by the other teams. The workshop facilitators also explained to the students various optical components and guided them through the whole setup -- most importantly, they reveal the location of the beam-splitter and the polarisation analyzer, which was not clandestinely placed yet often overlooked (Fig.~\ref{figure:student_setup}).

\subsection{Students feedback}\label{sec:feedback}

We conducted four runs of the workshop during school holidays in Jun 2018 and Jun 2019. Table~\ref{tab:response} shows the aggregated results from surveys conducted at the end of each workshop. 
Generally, students liked the workshop, and were able to follow and apply the concepts taught. 

\begin{table}[htb!]
    \small
    \centering
    \begin{tabular}{|p{15 cm}| p{2cm}|}
        \hline
        \hspace*{0.15cm} Reception (100\% responded)  & Average \\
        \hline
        \hspace*{0.3cm}1. How did you like the workshop? Range: 0 (Didn't like it) to 3 (It is great!) &  $2.67\pm0.57$\\
        
        \hspace*{0.3cm}2. How difficult did you find it? Range: -1 (Too easy!) to 1 (Too hard!) &  $0.04\pm0.33$\\
        \hline
        \hline
        \hspace*{0.15cm} What was your favourite aspect? (95\% responded) & Percentage\\
        \hline
        \hspace*{0.3cm}1. Successful eavesdropping attempt. & 31\% \\
        \hspace*{0.3cm}2. Hands-on experience to assemble and operate a functioning QKD system. & 29\% \\
        \hspace*{0.3cm}3. Opportunity to apply what they have learned in pre-workshop lectures. & 19\%\\
        \hspace*{0.3cm}4. Novelty of instructional method and worksheets. & 8\%\\
        \hspace*{0.3cm}5. Other comments. & 8\%\\
        \hline
        \hline
        \hspace*{0.15cm} What did you like the least? (54\% responded) & Percentage\\          
        \hline
        \hspace*{0.3cm}1. Lack of time available to experience the tasks of adjacent teams. & 12\% \\
        \hspace*{0.3cm}2. Waiting for facilitators to troubleshoot apparatus and programs. & 10\% \\
        \hspace*{0.3cm}3. Steep learning curve required to complete training materials and execute programs. & 10\%\\
        \hspace*{0.3cm}4. Insufficient clarity of instructions. & 5\%\\
        \hspace*{0.3cm}5. Other comments. & 17\%\\
        \hline
    \end{tabular}
    \caption{Workshop feedback aggregated from 59 pre-university students.}
    \label{tab:response}
\end{table}

\section{Conclusion}
We have presented a modified BB84 QKD setup aimed at pre-university students.
The setup was intentionally modified for ease of implementation while simultaneously providing a deeper insight into how the protocol critically relies 
on the underlying quantum physics - the no-cloning theorem.

The experiment is implemented with off-the-shelf, commercially available apparatus and can be operated with  minimal training in optics, computing and electronics. 
With the codes and documentations made available online, we believe that this simplified implementation of BB84 is an easily deployable and engaging QKD demonstration that can be realized by a wide range of educational institutions.

\section*{A. Encryption of Long Messages}

In the main text, we encrypted 32 bit messages with an XOR operation using the key generated from BB84 with the same length. 
However, for longer messages, we do not repeat the key generation procedure, but instead expand the 32 bit key to the same length as the message.
Although this procedure allows the students to encode the message with the XOR operation outlined above,~\cite{streamcipher} the resulting ciphertext is vulnerable to at least two different attacks.

First, we note that our key expansion procedure uses Mersenne Twister (MT) pseudo-random number generator (PRNG):~\cite{prng_mt, prng_python} the 32 bit key is used as an input (seed) to initialize the PRNG, deterministically producing a longer key. 
Given a partial knowledge of the key (``leaks''),~\cite{footnote_apdx1} an adversary can exploit the linear relations of the ``leaky'' bits and reconstruct the correct key.~\cite{mt_hack}
To prevent against this attack, one could implement a key expansion protocol using highly non-linear PRNG protocols.~\cite{cryptmt,othersc}

Second, an adversary can try all possible combinations of the 32 bit key, expand them, and decode the ciphertext with the XOR operation.  Performing this brute-force attack with computers, student would find out that only very few decoded messages would make sense.~\cite{footnote_apdx2}
It is important to note that every encryption protocol is prone to this type of attack, if the key is short enough.~\cite{cryptobook} 
The solution to this attack is to create a key long enough that requires an absurd amount of resources to be brute-forced (128 bits, with our current technology).

Although these vulnerabilities were not the main features of our workshop, we used them to highlight other potential weak links that could have been overlooked in a presumably secure cryptographic system. We observed that this was an interesting point for a few students, who attempted the brute-force attack by writing their own codes.

\section*{B. Cost and components}

The total cost of optomechanical, optical, and electronic components for the workshop was approximately USD 2.5k. However, as we were able to borrow most of the standard components from the optics lab and the electronics workshop, the out-of-pocket expenses were barely USD~450, most of which comes from purchasing Arduinos.
To reduce the cost, one can consider 3D-printing. 

The electrical components for the classical channel consist of a small solderless breadboard (MCBB4000), infrared LED (TSAL6200), infrared receiver (TSOP38238), bipolar NPN transistor (BC547), and various sizes of jumper wires and resistor.

The light source for the quantum channel was built with a reasonably-priced laser diode (Thorlabs L785P5) in a simple transistor switch circuit (see Fig.~\ref{figure:quantum_setup}). The light output was collimated with a lens (Thorlabs LT220P-B). Temperature and current stabilization were not necessary. 
The quarter wave plate (Dayoptics WPA215Q) was the most expensive optical component in our list, which cost around USD~100.
The motorized polarizers were built from a stepper motor (28BYJ-48-5V) driven with Darlington transistors (ULN2003) and a polarizer sheet (3DLens P50) glued to the motor shaft. The photodetection circuit consists of a PIN silicon photodiode (OP906) in a reverse-biased configuration, and a variable resistor for current-to-voltage conversion. 
The beam was aligned with high-reflective mirrors on kinematic mirror mounts (KM100).

\begin{acknowledgments}
We would like to thank Prof. Christian Kurtsiefer for his support and  guidance for this project, and for his valuable inputs on the manuscript.
We appreciate the help and support from Jenny Hogan and the steering committee of QCamp 2018 and 2019. 
We would like to extend our gratitude to all the students that took part in this experiment.
Last but not least, we gratefully acknowledge the help, effort, and perseverance from the following facilitators: Chow Chang Hoong, Tan Ting You, Yeo Xi Jie, Chen Jia Pern Neville, Florentin Thierry Adam and Ng Boon Loong.

This research is supported by the National Research Foundation, Prime Minister’s Office, Singapore and the Ministry of Education, Singapore under the Research Centres of Excellence programme.

\end{acknowledgments}


\begin{thebibliography}{99}
\bibitem{footnote1} The QWP converts linear-polarized light from the laser diode to circular-polarized light, which is subsequently projected into one of the four polarization states (H, V, D, A) with equal intensities using a motorized polarizer.
\bibitem{footnote2} Visibility measurement: First, Alice chooses an arbitrary rotation angle for her polarizer to fix the orientation of H-polarization. Then, she transmits a macroscopic beam with this polarization to Bob as a reference. To arrive on a common coordinate system with Alice, Bob rotates his polarizer and identifies the position that maximises the transmission of the reference beam with the H-polarization.
\bibitem{footnote3} To generate each polarization ``qubit'', Alice's microcontroller rotates her polarizer to the corresponding angle, and transmits through it a 200\,ms-long laser diode pulse. The overall duration required to generate each ``qubit'' (1.5\,s) is limited by our stepper motors (see Appendix B), which have a maximum rotation speed of about 10\,rpm.
\bibitem{footnote4} Synchronization procedure: Alice transmits a sequence of 16 ``qubits''. She precedes each transmission with a prearranged laser pulse sequence -- a header -- marking the beginning of each ``qubit'' sequence. Due to the clock-stability of the microcontroller ($\sim10^{-4}$), the uncertainty of the transmission time for each ``qubit'' is less than 2.4\,ms. The uncertainty is small compared to the duration of each ``qubit'', allowing Bob to predict the arrival times of the ``qubits'' after receiving the header pulse sequence.
\bibitem{footnote_extra} We aim to perform the key comparison step in future iterations of the setup, as it allows students to verify that their key was not corrupted by Eve using an intercept-resend attack. 
In our workshop, Eve intercepts, but does not resend ``qubits'', rendering her eavesdropping attempt immune to detection by the key comparison step.
Consequently, the impact of her success will be greater on the students who may have considered the key to be private given that they have performed key comparison but might have forgotten about the role that single quanta plays in the security of the protocol.
\bibitem{footnote5} Each transmission sequence is 16 ``qubits'' long and generates 8-bit $K$ on average. Thus, we repeated the sequence about four times to obtain a key 32-bit long.
This process takes about 2.5\,minutes.
\bibitem{footnote6} Eve could have performed the polarization calibration procedure along with Alice and Bob, which allows her to select the best polarization measurement for distinguishing and identifying the intercepted polarization states. However, given the limited time for the workshop, it was difficult to prepare the students in team Eve in time for the calibration procedure as they would then have to learn calibration, principles of the quantum channel, using two polarization measurements to extract the polarization from the intercepted photons, and the clustering algorithm - the latter two concepts that Alice and Bob did not have to learn.
\bibitem{footnote7} There are several eavesdropping noise sources. First, the laser intensity fluctuates ($\sim 20\%$) with the position of the polarizer sheet, which is sensitive to mechanical disturbances (e.g. wind, students' movements). Moreover, the current to the laser diode was also not feedback-stabilized. These noise sources affect the result if the measurement is only performed in one basis (see e.g. Photodiode 1 histogram in Fig.~\ref{figure:kmeans}). To optimally distinguish between the various polarization states, Eve projects each state in two measurement basis, each separated by about 45$^{\circ}$ from each other, and subsequently measure the corresponding intensity.
\bibitem{footnote_apdx1} The ``leaks'' of the expanded key can, for example, be obtained by examining the structure of the message. For the ASCII or UTF-8 schemes which are widely used to represent characters in a message, the two leading bits of an English alphabet is 01. 
This could be used to deduce the leading bits of every byte of the expanded key by sampling the ciphertext.
\bibitem{footnote_apdx2} For any message length $m$, this operation results in only $2^{32}$ possible combinations of decoded messages. If the message is much longer than the key ($ m \gg 32$), it is likely that only a few decoded messages are coherent. This contrasts with the scenario where the message is as long as the key ($m = 32$, one-time pad scenario), as this results in as many coherently decoded messages as possible. Students can perform this brute-force attack with modern computers -- if each combination requires $\sim1\,\mathrm{\mu s}$ to be computed and checked for coherence, it would only take $\sim1.2$ hours to exhaust all possibilities.
\end{thebibliography}

\end{document}